\documentclass[letterpaper,11pt]{article}
\usepackage[letterpaper, left=.8in, top=.8in, right=.8in, bottom=.8in,nohead,includefoot]{geometry}
\usepackage{charter, bm}
\RequirePackage{authblk,natbib}

\usepackage[hidelinks]{hyperref}
\usepackage{xcolor}

\begin{document}
 
\begin{center} 
\large
Reply to Discussions of 

\medskip
{\bf Multivariate Dynamic Modeling for Bayesian
Forecasting of Business Revenue}

\medskip\large Yanchenko, Tierney, Lawson, Hellmayr, Cron \& West \medskip

\normalsize \today
\end{center}

We are most grateful to all discussants for their positive comments and many thought-provoking questions to our paper (\citealp{Yanchenko:2022}). In addition, the discussants provide  a number of useful leads into various areas of the literatures on time series, forecasting and commercial application within which the work in our paper is, of course,  just one contribution linked to multiple threads. Our view is that, collectively, the discussion contributions  nicely expand on the core of the paper and together-- with multiple additional references-- provide an excellent point-of-entr\'ee to the broader field of retail forecasting and its research challenges. Interested readers are encouraged to dig deeply into the discussions and our responses here, and explore referenced sources. 

There are several themes that recur across discussants, as well as a range of specific  points/questions raised.  Following some \lq\lq big-picture'' comments on our perspectives on Bayesian forecasting systems, we comment in turn on some specifics in each contribution.

\subsection*{Perspectives} 

Central, cross-cutting perspectives on Bayesian forecasting  are critically relevant to some of the points raised. These include:  addressing and dealing with unexpected events and changes over time beyond that described in a particular model, or set of models; related questions of adapting to unforeseen or \lq\lq rare'' events; questions of integrating information from multiple sources into formal forecasting models-- including information from other, related models or models at different levels of time resolution that draw on different data sets, human inputs and subjective opinions, and varieties of partial constraints; and, critically, the roles of interpretable modeling in connection with these (and other) challenges.  The Bayesian forecasting philosophy has, for decades, been that of integrating models  with decision makers and contextual constraints in the organization-- as part of the overall forecasting \lq\lq system''. Within this, we stress the role of methods for formal model/forecast monitoring that are open to  adapting to the impact of environmental, economic, commercial and other changes. This stresses the importance of relevant methodology for formal  adaptation of models in response. The latter includes but is not limited to the use and 
 integration of multiple forms of external information into a  forecasting model.  The roles of  subjective model adjustments as well as decision-guided automatic interventions have been foremost in the Bayesian forecasting community~(e.g.~\citealp{West1997}, chapters 1 and 11 and references therein; \citealp{West1989}; \citealp{West2023constrainedforecasting}). This broad perspective on the contributions of formal modeling, and the needs for models to be open and responsive to many kinds of end-user interests as well as multiple forms of potential interventions in sequential evolution over time, is  fundamental to  operational forecasting. This is true for forecasting in any field, but perhaps especially so in retail enterprises and allied commercial settings.

\subsection*{Mike K.P. So and Cathy W.S. Chen}
The discussants highlight three main areas for further development: (i) addressing questions of scale, (ii) looking at block multi-scale structure across LSGs and/or product categories, and (iii) adapting to rare or unforeseen events (e.g., the COVID-19 pandemic; \citealp{So:2023}). These are important areas of focus for 
the current setting of retail forecasting, as they are in other application areas. 

The question of scaling to larger numbers of LSGs and product categories is partly addressed within  the overall decouple-recouple framework of our dynamic models.  As $c$ and $z$ grow, it is possible to model, at the first step, $Y_{t, c} | X_{t, c}$ in parallel across all LSGs, $c$. This can enable scaling to larger numbers of LSGs.  After the multi-scale discount information $m_{t,c}$ is extracted, modeling can proceed in parallel across product categories in the second step defining $Y_{t, c, z} |  X_{t, c, z}, m_{t, c}$.  In addition to computational questions, Professors So and Chen  raise good points about the utility of the multi-scale signal $m_{t, c}$ as the number of LSGs grows.  It is likely that a subset of all LSGs inform most directly on specific product categories, suggesting a block multi-scale structure.

Utilizing a block multi-scale structure is certainly a good idea and offers an approach to address scaling. Such an approach may also improve forecasting by restricting sharing of information to relevant LSGs and product categories;  see the brief comments on, for example, the alcohol category in the Summary Comments of the main paper.  In our retail setting, in particular, it is likely that clear blocks of LSGs and product categories can be formed based on external information about the geographic regions of stores, the hierarchy of products, etc.  It also may be possible to assume that these blocks are relatively static over time, making inference easier as the discussants mention. 

There are, however, other applied settings where there might not be clear block structure that is known {\em a priori}, and/or in which relevant block structure is stochastic and time-varying, again as noted by the discussants. This will introduce additional modeling challenges, especially as multiple  different block formats will likely yield similar predictive performance depending on the application~\citep[e.g.][]{West2020Akaike}. While predictive performance alone can be used to tune and select  block structure, a general approach that also integrates domain knowledge is preferable.  

In terms of adapting the underlying dynamic models to rare events, such as the COVID-19 pandemic,  we refer to our lead-in commentary on model interpretability, monitoring, intervention and related matters in our {\bf Perspectives} section at the start of this response to the discussions. 
In applied settings such as ours, it can be challenging to predict how such events will differentially impact different LSGs and product categories.  For example, in the case of some rare events it is likely that any imposed block multi-scale structure would need to  adapt.   This key point of the discussants relates to the need for further exploration in large, interdependent systems of how best to intervene for these types of events.

\subsection*{Chris Glynn, Ida B. Johnsson and Mingzi Yi}

The discussants provide detailed and broad-ranging commentary from a core applied perspective; they touch on multiple important points 
about the opportunities and challenges of statistical model-based forecasting in an enterprise setting.  Three highlighted challenges   are (i) scaling to enterprise-wide data settings, (ii) sharing information across both products and areas of the business, and (iii) framing and communicating forecast information such that stakeholders across the enterprise can interpret and utilize the results.   The discussants note that  statistical models themselves are typically one small part of the larger forecasting ecosystem across a business; expanding on our approaches to facilitate easier design, testing, distribution and communication of results to various stakeholders is critical.

We concur with the discussants that-- in terms of specific technical areas for development--  multi-scale Bayesian dynamic models are particularly well-suited to addressing many of these real-world forecasting challenges.   The interpretability of  dynamic models supports communication of results to many different types of teams across the enterprise. Coupled with the full probabilistic representation of the forecasting system, this opens the path to intervention as needed.  Then, full probabilistic forecasts allow for various loss functions to be exploited for different end-uses, and enables exploration of  ``What-if?'' scenarios of interest.   Finally, the general multi-scale approach enables sharing of information across especially relevant hierarchies or structures in the business. 

The underlying DLMs are   flexible and  open to accommodating modifications based on various sources of external information.  We refer to our lead-in commentary on model interpretability, monitoring, intervention and related matters in our {\bf Perspectives} section at the start of this response.   Then, as the discussants note, it can be challenging to accommodate data, such as macroeconomic data, that arrive at different times and frequencies.  One possible solution is to further extend the multi-scale structure of the models across temporal resolutions, to enable sharing of information across the time scales of interest. See also our comments in response to Korobilis and Montoya-Bland\'{o}n  for related discussion.

\subsection*{Dimitris Korobilis and Santiago Montoya-Bland\'{o}n}

The discussants have a number of relevant comments and suggestions for extensions to the current work (\citealp{Korobilis:2023}). We concur with their views about the importance and flexibility of Bayesian dynamic modeling for accommodating many real-world forecasting challenges, especially when in a multi-scale framework in which information sharing is appropriate and can be key to improved forecasting.  Three specific areas highlighted by the discussants are 
those of (i) incorporating additional predictors and dealing with mixed-frequency data, (ii) use of external information, and (iii) choices of loss functions. 

In terms of exploring the utility of additional predictors and other model interventions (such as the use of dynamic discount factors), we agree  that  forecasting performance may be improved with additional such measures, some of which may be more or less relevant at different periods of time.    Bayesian forecasting models as part of a forecasting system are not, however  {\em\lq\lq updated solely by information in the data''}. 
The Bayesian forecasting literature has always stressed the importance of routine openness of models to interventions over time, and the use of dynamic discount factors has always been part of this, preceding the referenced work the discussants add to the broader picture~\citep[e.g.][]{West1986a,West1986}.  We also refer to our lead-in commentary on model interpretability, monitoring, intervention and related matters in our {\bf Perspectives} section at the start of this response to the discussions.  

We   appreciate the links to recent work on dynamic variable selection, a research area that has expanded substantially in recent years (see, for example, discussion and literature cited in~\citealp{West2020Akaike}).   Our own work on this has been partially represented by two related approaches. First, the foundational approach of {\em dynamic latent thresholding modeling}, explicitly focused on the questions of time-adaptivity in the roles of predictors and their forecasting implications~\citep[e.g.][]{NakajimaWest2013JBES,NakajimaWest2013JFE}; second, on the approach using SGDLMs~\citep{GruberWest2016BA,GruberWest2017ECOSTA}  in which the philosophy is to stick with a given model, or set of models, unless the forecasting performance degrades or can likely be improved by switching some predictors in/out, again adaptively over time. 
 
We have limited experience with the use of mixed-frequency data, but recognize the importance of   recent research in, particularly, the Bayesian econometrics literature.  As noted by these discussants, as well as by  Glynn, Johnsson, and Yi,  it is critical in many real-world forecasting settings to incorporate mixed frequency information, such as macroeconomic data, into forecasting models at a more \lq\lq micro'' scale such as in the retail forecasting setting.  While Bayesian dynamic models are naturally open to  incorporating such information-- either by treating some values as missing as the discussants note, or by exploring temporal multi-scale structure-- this is a key and open area for near-term research.  We also note the connections with research on imposing constraints (especially partial stochastic constraints) on forecast distributions from a given model; advances in methodology for coherent analysis within a Bayesian framework will be relevant to that~\citep[e.g.][]{TallmanWestET2022,West2023constrainedforecasting}. 
  
The suggestion related to accounting for various cross-correlation structures  resonates with suggestions of discussants So and Chen, and our responses detailed above on various choices of block multi-scale structure.  Appropriately choosing the multi-scale structure is likely to   improve model scalability and  predictive accuracy.  We agree that approaches as in \citet{GruberWest2016BA, GruberWest2017ECOSTA} provide some basis for  more formal Bayesian methods for identifying predictively relevant block structures and their changes over time.

We also agree that choosing an appropriate loss function is critical in producing relevant point forecasts and is   dependent on the application goals. While MAPE is commonly used in the retail forecasting setting, different settings necessitate other loss functions, which again are easily supported by the probabilistic forecasts produced by DLMs.  We note the use of multiple different loss functions in some of our prior work on retail forecasting~\citep{BerryWest2018DCMM,BerryWest2018TSM} including the introduction of new classes of ZAPE loss functions for contexts involving forecasting non-negative integer time series. We also stress a broader decision analysis view that  explicitly includes investigation of full predictive uncertainties of loss-- rather that just following the typical  \lq\lq act on the optimal decision''-- that can highlight potentially important practical considerations~\citep[e.g.][]{West2023constrainedforecasting}. 
 
\subsection*{Dawid Bernaciak, Jim Griffin and Ioanna Manolopoulou}

The discussants provide insightful and detailed comments on  technical and applied aspects of our work, and particularly useful discussion on
retail forecasting challenges with additional references to intersecting literatures (\citealp{Bernaciak:2023}).  This discussion should be regarded as core reading for newcomers to the areas.    Then, their suggestions focus on three main areas of further methodological extensions: (i) questions related to aggregation, (ii) demand modeling, and (iii) full uncertainty quantification.  
 
We agree  that the practical impacts of aggregation across LSGs depend strongly on the LSGs under consideration and the heterogeneity across LSGs.  As mentioned by discussants So and Chen,  and by Korobilis and Montoya-Bland\'{o}n, one approach to addressing and potentially mitigating this concern is to model subsets or blocks of LSGs and to only share information across relevant dimensions. A similar approach of block multi-scale structuring could also be used to enable  heterogeneity in the predictors by LSG and product category, as the discussants suggest.  

In our retail revenue setting, we see high dependence between  revenue and demand, and thus chose to model revenue directly. We do agree, however,   that there are many settings where modeling demand, especially basket-level demand, is of primary interest.  One approach to modeling basket-level demand in a decouple/recouple framework was a core advance in some of our prior work on supermarket sales forecasting.  This developed a coupled approach to forecasting transactions and sales via novel {\em dynamic binary cascade models}, or DBCMs~\citep{BerryWest2018TSM} with major potential improvements in forecast accuracy induced. We do  agree with the discussants that there are many opportunities for extending multi-scale modeling concepts to basket-level demand. We also, however, add the caveats about understanding limitations of \lq\lq demand data'' due to censoring based on product shelf availability and other constraints. The discussants also raise the important point that, in addition to improving forecast accuracy, sharing information in the hierarchical, multi-scale setting can enable modeling of new products for which no historical data is available. 
 
We agree that the choice of loss function is specific to the application of interest; see discussion of Korobilis and Montoya-Bland\'{o}n, and our comments in response. In particular,  our prior work on retail forecasting~\citep{BerryWest2018DCMM,BerryWest2018TSM} introduced new loss functions 
specifically customized to the retail forecasting challenges at very low levels of demand/sales for many items.  Analyses then evaluated forecasting models using multiple loss functions.  Decision analysis that  incorporates  full forecast uncertainty is especially relevant in many business settings, in particular when evaluating the impact of various modeling choices on down-stream decision making.   As noted by Glynn, Johnsson, and Yi, and by Yelland, forecasts are often used by many departments across the enterprise system, and the form of decisions varies by department.  Including  full forecast uncertainty can significantly expand perspectives on the applied impact of decision analysis that recognizes second-level uncertainty beyond the normal analysis that conditions on optimal forecast decisions.  As noted in response to Korobilis and Montoya-Bland\'{o}n, we also stress the explicit study of full predictive uncertainties of loss as a general principle in Bayesian decision analysis~\citep[e.g.][]{West2023constrainedforecasting}. 
Finally, on our technical use of \lq\lq plug-in'' multi-scale forecasts as a computational short-cut of the full Bayesian analysis,  evaluation across various metrics and decision-making settings could indeed be useful. 

Dealing with significant holiday effects can be challenging and have long-term impacts due to the 12-week forecasting horizon.  Ideally, with more historical data these effects can be addressed explicitly with appropriate seasonality in  DLMs.  In  more limited data settings, holiday effects provide another example of the importance of incorporating external information. There are various technical strategies to use such intervention information~\citep[e.g.][chapters 1 and 11 and references therein]{West1997}. We refer to our lead-in commentary on model interpretability, monitoring, intervention and related matters in our {\bf Perspectives} section at the start of this response to the discussions.

\subsection*{Graham Sparrow}

We much appreciate the commentary on the roles of models, the questions of model interpretability, and importance of probabilistic framing to properly characterize forecast uncertainties for potential various end-users and decision makers (\citealp{Sparrow:2023}).  The importance of these aspects is highlighted in this 
discussion from a core applied perspective in commercial forecasting.   Our response to the discussion of 
Bernaciak,  Griffin and Manolopoulou immediately above touches specifically on the latter point-- that of uncertainty quantification and propagation through to decision analysis, as well as the broader questions of model interpretability and the intersection with open intervention as part of a broader forecasting system. 

On some detailed points, the raised comment about exploring which specific categories would most benefit from the inclusion of multi-scale information-- and if there are specific points in time when multi-scale models would be most effective-- relates to the structure of multi-scale dependence in any chosen model.  One method to tie features of the categories to potential benefit from multi-scale information would be to consider a block multi-scale structure, as noted in the discussion contribution of So and Chen,  and to relate the determined or learned hierarchy to features of the various product categories.

The interpretability of DLMs is critical to the business setting of interest here, and the discussant raises an  interesting point about exploring how significantly various coefficients of interest need to change for this change to lead to practical business significance. One way of approaching this question could be as suggested above by Bernaciak, Griffin, and Manolopoulou-- in tying various evaluation metrics to actual impact on business decisions. 
We again also refer to the relevance of our lead-in commentary on model interpretability, monitoring, intervention and related matters in our {\bf Perspectives} section at the start of this response to the discussions.  

The discussant raises-- almost in passing-- the major and critical issue of causality.  We have not overtly addressed this in the current paper, but it emerges, of course, as a key and central question in decision analysis in retail forecasting as in other areas (including macro-economic forecasting in policy settings, for example).  The discussant asks about potential for  identifying complementary and substitution effects in retail sales,  and explicitly from a causal perspective. While not addressing that specific goal, some of our recent, related work is wholly focused on causal inference in sequential forecasting with Bayesian dynamic models,  emphasizing multivariate dynamic models for time-varying effects across multiple treated units  and sequential learning of effects of interventions~\citep{TierneyEtAl2023}. The approach, extending the time series literature on use of multiple potential synthetic control variables in a multivariate setting,  is developed in a study of interventions in a supermarket promotions experiment. We regard this as an approach to expand across retail forecasting in areas such as raised by the discussant.  
 
\subsection*{Phillip Yelland}
  
The discussant presents a detailed commentary on the \lq\lq untidier'' aspects of retail forecasting, emphasizing the the need for comprehensive exploration of the roles of formal statistical models within the multi-faceted environment of modern retail enterprises. We particularly appreciate this broad perspective from an experienced Bayesian forecaster who has driven forecasting developments in multiple large-scale companies.  Yelland lays out the panoply of considerations faced by forecasting groups and industry leaders in large-scale commercial environments (and goes well beyond supermarket systems and allied retail organizations).  This discussion, and the short recent article by the author~\citep{Yelland2019}, should be recommended reading for newcomers to business forecasting and researchers interested in engaging in R\&D in these areas.   We are grateful to the discussant for the commentary and sharing perspectives, and have little to add but endorsement of the main points.  To some degree,  these relate to our lead-in commentary on model interpretability, monitoring, intervention and related matters in our {\bf Perspectives} section at the start of this response to the discussions; and then go beyond to reflect on the realities and consequent challenges of using statistical forecasting models in practice in the enterprise.  Many of these points overlap with those raised by other responses above. 

On some specific points, we note the questions of (i) forecast (system) reconciliation, (ii) unreliable covariates (\lq\lq signals'') and the dynamics and uncertainty in promotion campaigns (and intervention \lq\lq experiments'' more generally), with  linked questions of causality, 
and (iii) post-hoc forecast adjustments and broader issues of user manipulation/modification of outputs of formal forecast models.    Our prior work bearing on forecast (system) reconciliation has been touched on above, 
including in response to Korobilis and Montoya-Bland\'{o}n, in connection with forecast model combination and imposing constraints on a given model informed by external information (or forecasts from other models;~\citealp[e.g.][and references therein]{TallmanWestET2022,West2023constrainedforecasting}).     These comments also connect this stream of methodology to the integration into down-stream decision making by potentially multiple end-users, and the developing literature on forecast model-external data synthesis. 

The challenges of \lq\lq noisy covariates'' is a central question in much of statistical forecasting, and is particularly acute here in connection with the inherently causal interests in evaluating price/promotion and other incentive strategies in influencing consumer behavior.  Appropriately characterizing and quantifying the realized value of promotion campaigns-- which are inevitably subject to multiple sources of stochastic influence in terms of how they are implemented, and with little or no real understanding of how they are perceived by customers-- is among the major challenges in this area and one main area of our current interest in developing causally focused analyses~\citep{TierneyEtAl2023}.  In this area, rolling sets of promotional activities often interact over time, and are played out at different time periods in different stores or regions, or with respect to different sections of the consumer base. Teasing out relevant and 
\lq\lq reliable'' intervention covariates from the resulting highly \lq\lq complex, dynamic'' signals is surely a major priority across industries, and a central research area for statistics and stochastic modelers generally.

At a higher-level, while we agree that Bayesian DLM approaches are amenable to development to address all such (and other) challenges,  specific, stylized extensions will need to be context-specific.  Each of these areas represents open questions in need of basic and methodological research, as well as integration of that research into use-cases to drive conceptual and technical innovation in practically relevant directions.  
 
\break\newpage

\bibliographystyle{chicago} 
\bibliography{ASMBI-response}

\end{document}